\documentclass{article}
\usepackage{amssymb,amsfonts,amsmath, psfrag,eepic, epsfig, graphicx}

\title{Some Numerical Results For  Ito Equation}

\author{\\
YuQi Li $^a$ \footnote{E-mail: liyuqi@nbu.edu.cn}, 
Biao Li $^a$ \footnote{E-mail: biaolee2000@yahoo.com.cn}\\
\small $a$
Center for Nonlinear Science, Ningbo University\\
Ningbo 315211, China}

\date{}
\begin{document}
\maketitle

\footnote{The work
is partly supported by  NSF of China(No.11041003),
NSF of Ningbo City (2009B21003, 2010A610103, 2010A610095).}

\begin{abstract}
By the method of invariant manifold, we investigate
the Ito equation  numerically with high precision.
By the numerical results, we can completely determine the form of
analytic soliton solutions for the Ito equation.
In fact, by the numerical data 
we have succeeded in deciding the analytic form of 
the $\tau$-function, which is more general than the old assumptions. 
This may suggest that 
we should think more deeply about what the soliton is.

keywords: Ito equation, invariant manifold, $\tau$-function.
\end{abstract}

\section{Introduction}
\hspace*{0.6cm}The Ito equation \cite{Ito} was first introduced in 1980
as the Hirota bilinear equation
\begin{eqnarray}
D_t (D_t+D_x^3) f\cdot f=0. \label{ItoBilinear}
\end{eqnarray}
By introducing variable $u=2 (\ln f)_{xx}$,
Equation (\ref{ItoBilinear}) is transformed to a soliton equation
\begin{eqnarray}
u_{tt}+u_{xxxt}+3 (2 u_x u_t+u u_{xt})+3 u_{xx} \int_{-\infty}^x u_t dx=0.
\label{ItoIntForm}
\end{eqnarray}

Non-local equation (\ref{ItoIntForm}) is usually written into a local one
by the transformation
$u(x,t)=\tilde u(x,-t)$,
$\frac{\partial}{\partial t}{\tilde u(x,t)}=
3 \frac{\partial}{\partial x}  {\tilde v(x,t)}$.
Neglecting the tilde symbol and the integral constant, we get
\begin{eqnarray}
&&u_t=3 v_x, \nonumber\\
&&v_t=(u v)_x+v_{xxx}. \label{ItoLoalForm}
\end{eqnarray}
In the sequel system (\ref{ItoLoalForm}) will be referred to as Ito equation. 
In Ref.\cite{Drinfeld-jsm-1985}, 
Drinfeld and Sokolov also found the equation (\ref{ItoLoalForm}) 
in their study on the relationship between soliton equation of KdV type 
and Lie algebras, thus some authors referred (\ref{ItoLoalForm}) 
as Drinfeld-Sokolov system \cite{gurses-jmp-1999}. 
The Hamiltonian structures of Ito equation was obtained 
by Liu \cite{liuqp-pla-2000}.

The Lax pair  for (\ref{ItoLoalForm}) is \cite{Drinfeld-jsm-1985}
\begin{eqnarray}
&&\hat L=\partial^3+u \partial +u_x+v \partial^{-1}, \nonumber\\
&&\hat P=\partial(\partial^2 +u). \label{ItoLax}
\end{eqnarray}

Also in  \cite{Ito} Ito himself gave a  solution of (\ref{ItoBilinear})
\begin{eqnarray}
f=\sum_{\mu=0,1} \exp\left(
\sum_{i<j}^{(N)} C_{ij} \mu_i \mu_j +\sum_{i=1}^N \mu_i \eta_i
\right), \label{ItoSolF}
\end{eqnarray}
where
\begin{eqnarray}
&&\eta_i=p_i x+\Omega_i t+\eta_i^0, \nonumber\\
&&\Omega_i=-p_i^3, \nonumber\\
&&\exp C_{ij}=\frac{(p_i-p_j)(p_i^3-p_j^3)}{(p_i+p_j)(p_i^3+p_j^3)}.
\label{ItoParaVal}
\end{eqnarray}

It had been a  long time taking  (\ref{ItoSolF}) as the
$\tau$-function of the soliton solution of (\ref{ItoIntForm}).
But one has good reason to question
whether (\ref{ItoSolF}) covers
all soliton $\tau$-function of the Ito equation:
literature \cite{Ito} has assumed that
for soliton solutions $f$ is a sum of exponential traveling waves.
Zhang and Chen \cite{ZhangY}  are among the first ones to
try to enlarge the class of soliton solutions of the Ito equation.
It seems to us that the solutions obtained in  \cite{ZhangY}
are still too limited.
Li et al.  \cite{LLL} had investigated the Ito equation by
the method of invariant manifold (IM) and
solved  the first two constraints completely.
But they \cite{LLL} did not give the general solution
for the higher order constraints.
The results of this paper will reveal some important structures
for  higher order constraints discussed in \cite{LLL}.

This paper is organized as follows.
Section 2  will derive the ordinary differential equations (ODEs)
for solving the IMs suggested by \cite{LLL}.
The ODEs in Ref.\cite{LLL} is not suitable for numerical integration. 
Here we  give the suitable one.
Also here we add an equation for $\mathsf{f}$,
which is crucial to analyze the IMs.
Section 3 will present the numerical results
for the ODEs obtained in section 2.
These numerical results will reveal the general structure of $\tau$ function
corresponding to the 2-soliton and 3-soliton solution of the Ito equation.
Section 4 is the conclusions.

\section{ODEs for the time evolution of
solitons of the Ito equation}

For brevity we use the following notations:
$f'=\frac{\partial f}{\partial x}$,
$\dot f=\frac{\partial f}{\partial t}$.
The definition of  $\partial^{-1}$  is
$\partial^{-1} \partial=\partial \partial^{-1}=1$ \cite{Dickey}.

With eigenfunction $\psi_i$,
the Lax pair (\ref{ItoLax}) becomes $\hat L \psi_i=\lambda_i \psi_i$ and
$\dot \psi_i =\hat P \psi_i$,
which are equivalent to the following differential equations
\begin{eqnarray}
&&\left(\frac{\lambda_i \psi_i -\psi_i'''-(u \psi_i)'}{v} \right)'=\psi_i
\nonumber\\
&&\dot \psi_i=\psi_i''' +(u \psi_i)'. \label{ItoLaxPsi}
\end{eqnarray}
According to \cite{LLL}, the constraint for Equation (\ref{ItoLaxPsi}) is
\begin{eqnarray}
v=\sum_{i=1}^n a_i \psi_i, \label{ConstrV}
\end{eqnarray}
where $a_i$s are real or complex numbers and
we can take $a_i=1$ without loss of generality.

Let $\psi_i=\varphi_i'$. Then (\ref{ItoLaxPsi}) becomes
\begin{eqnarray}
&&\lambda_i \varphi_i'-\varphi_i''''-(u \varphi_i')'=v (\varphi_i+k_i(t)),
\nonumber\\
&&\dot \varphi_i=\varphi_i'''+u \varphi_i'+c_i(t).\label{ItoLaxVPsi}
\end{eqnarray}
It is easy to prove $k_i(t)=0$ and $c_i(t)=0$.

Altogether we have
\begin{eqnarray}
&&\lambda_i \varphi_i'-\varphi_i''''-(u \varphi_i')'=v \varphi_i,
\label{LaxL}\\
&&\dot \varphi_i=\varphi_i'''+u \varphi_i',\label{LaxP}\\
&&v=\sum_{i=1}^n \varphi_i', \label{ConstrVv}\\
&&\dot u=3 v'. \label{EvolU}
\end{eqnarray}

{\bf Remark 1:} From Equations (\ref{LaxL}) and (\ref{ConstrVv}), 
we can easily  see that with respect to $x$ 
there are $n$ equations with $n+1$ unknown functions: 
$\{\varphi_1,\varphi_2,\cdots, \varphi_n, u\}$. 
So we can select $u(x,t)|_{t=0}=u_0(x)$ to be an arbitrary function, 
i.e.,  with respect to $x$,
$u$ is not limited to the form $2 (\ln \sum a_i e^{k_i x})_{xx}$ \cite{Ito}.

By (\ref{LaxL}),  (\ref{LaxP}),  (\ref{ConstrVv}) and  (\ref{EvolU}),
the time evolution of
$u$, $\varphi_i$, $\varphi_i'$, $\varphi_i''$, $\varphi_i'''$
become a closed system of ODEs:

\begin{eqnarray}
\dot u &=&3 v'. \nonumber\\
\dot \varphi_i &=&\varphi_i'''+ u \varphi_i'. \nonumber\\
\dot {\varphi_i'}&=&(\varphi_i'''+ u \varphi_i')' \nonumber\\
                 &=&\lambda_i \varphi_i'-v \varphi_i, \nonumber\\
\dot {\varphi_i''}&=&(\lambda_i \varphi_i'-v \varphi_i)' \nonumber\\
                 &=&\lambda_i \varphi_i''-v' \varphi_i-v \varphi_i',
\nonumber\\
\dot {\varphi_i'''}&=&( \lambda_i \varphi_i''-v' \varphi_i-v \varphi_i')'
 \nonumber\\
   &=&\lambda_i \varphi_i'''-v'' \varphi_i-2 v' \varphi_i'-v \varphi_i'',
\label{ODEs}
\end{eqnarray}
where $i=1,2,3,\cdots,n$.
Clearly the ODE system (\ref{ODEs}) contains $4 n+1$ equations.

Next we will introduce an extra differential equation
\begin{eqnarray}
\dot {\mathsf{f} }=\sum_{i=1}^n \varphi_i. \label{Evolf}
\end{eqnarray}
Equation (\ref{Evolf}) is very important
to reveal the solution  structure of  Equations (\ref{ODEs}).

Summing (\ref{LaxL}) over $i$, we immediately get
\begin{eqnarray}
\sum_{i=1}^n (\lambda_i \varphi_i'-\varphi_i''''-(u \varphi_i')')
=\frac{1}{2} \left( \left( \sum_{i=1}^n \varphi_i \right)^2 \right)' .
\label{PreU}
\end{eqnarray}
Integrating (\ref{PreU}) we get
\begin{eqnarray}
\gamma=-\frac{1}{2}  \left( \sum_{i=1}^n \varphi_i \right)^2
+\sum_{i=1}^n (\lambda_i \varphi_i-\varphi_i'''-u \varphi_i').
\label{gammaDef}
\end{eqnarray}
It is easy to prove $\gamma$ is independent on $t$.
We will use the conservation of $\gamma$
to illustrate the precision of the numerical solutions.

\section{Numerical results}
Because the case $n=1$ had been completely solved in \cite{LLL},
we start numerical study from $n=2$.

\subsection{n=2}
The parameters $\lambda_i$, $i=1,2$,
can be taken any value in Equations (\ref{ODEs}).
For simplicity we take
\begin{eqnarray}
\lambda_1=1, \quad \lambda_2=2 .\nonumber
\end{eqnarray}

The initial values for unknown functions in (\ref{ODEs}) and (\ref{Evolf})
are also arbitrary.
In order to make the numerical solving procedure simple
we randomly choose them as
\begin{eqnarray}
&&\varphi_1(0)=1, \quad \varphi_2(0)=1, \nonumber\\
&&\varphi_1'(0)=1, \quad \varphi_2'(0)=-1, \nonumber\\
&&\varphi_1''(0)=1, \quad \varphi_2''(0)=1, \nonumber\\
&&\varphi_1'''(0)=2, \quad \varphi_2'''(0)=2,\nonumber\\
&&u(0)=-3,\quad  {\mathsf f}(0)=0. \label{Initials2}
\end{eqnarray}

\begin{center}
\begin{minipage}{11cm}
\begin{center} \includegraphics[width=10cm]{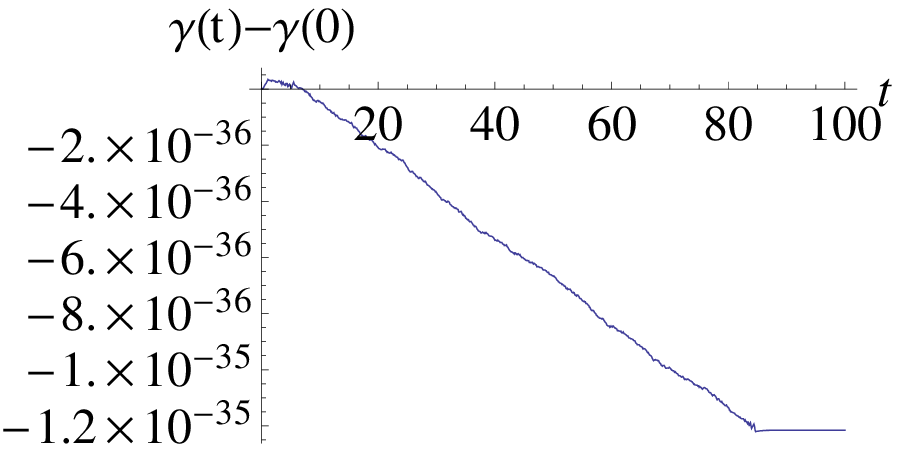} \end{center}
\begin{center}
Fig. 1 \quad The difference between $\gamma(t)$ and $\gamma(0)$.
\end{center}
\end{minipage}
\end{center}

The calculated $\gamma$ is very close to $\gamma(0)=-3$.
Fig. 1  shows  that the numerical precision in our numerical integration
for the ODEs (\ref{ODEs}) is very high.

Numerical results show that for lots of initials
$\varphi_i$s are asymptotically to constants.
For initial values (\ref{Initials2}), the numerical results are
\begin{eqnarray}
&&\varphi_1(+\infty) \approx \varphi_1(100)=
-8.515821092114916270761375823327277321106, \nonumber\\
&&\varphi_2(+\infty) \approx \varphi_2(100)=
15.71501739387443269240781890998386797269. \nonumber\\
&&\label{vpsis2Inf}
\end{eqnarray}

\begin{center}
\begin{minipage}{11cm}
\begin{center} \includegraphics[width=10cm]{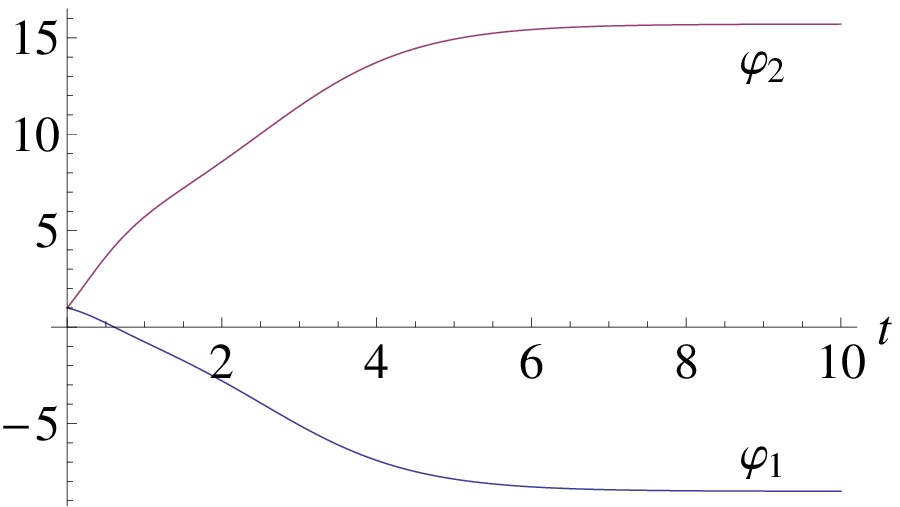} \end{center}
\begin{center}
Fig. 2 \quad The plot of $\varphi_1$  and $\varphi_2$.
\end{center}
\end{minipage}
\end{center}

Fig. 2 shows that $\varphi_1$ and $\varphi_2$ will approach to $\varphi_1(+\infty)$
and $\varphi_2(+\infty)$ respectively.

By the third equation of (\ref{ODEs}),
we can show that the asymptotic behaviors of $\varphi_1'$
and $\varphi_2'$ are determined by
\begin{eqnarray}
M=\left( \begin{array}{cc}
\lambda_1-\varphi_1(+\infty)&-\varphi_1(+\infty)\\
-\varphi_2(+\infty) & \lambda_2-\varphi_2(+\infty)
\end{array} \right). \label{DefM2}
\end{eqnarray}

The eigenvalues of $-M$ are
\begin{eqnarray}
&&\Omega_1=3.1444367053092452142565168204216716495466, \nonumber\\
&&\Omega_2=1.0547595964502712073899262662349190020374.
\label{EigenM2}
\end{eqnarray}

Define
\begin{eqnarray}
\omega_1=\frac{1}{2} \Omega_1, \quad \omega_2=\frac{1}{2} \Omega_2.
\label{Defomega2}
\end{eqnarray}
By the numerical values of $\varphi_1(t)$ and  $\varphi_2(t)$,
 $t \in (0, 3)$,
we can determine
\begin{eqnarray}
\varphi_1(t)=\frac{\sum \limits_{i,j \in \left\{1,-1 \right\} }
a^1_{i,j}e^{i \omega_1 t+ j \omega_2 t}}
{\sum \limits_{i,j \in \left\{1,-1 \right\} }
b_{i,j}e^{i \omega_1 t+ j \omega_2 t}},
 \quad
\varphi_2(t)=\frac{\sum \limits_{i,j \in \left\{1,-1 \right\} }
a^2_{i,j}e^{i \omega_1 t+ j \omega_2 t}}
{\sum \limits_{i,j \in \left\{1,-1 \right\} }
b_{i,j}e^{i \omega_1 t+ j \omega_2 t}},
\label{varphiForm2}
\end{eqnarray}
where we can assume $b_{-1,-1}=1$.
Note that ($\ref{varphiForm2}$) are obtained  by only a fuzzy thinking.
The calculated $a^1_{i,j}$,  $a^2_{i,j}$ and $b_{i,j}$ are
\begin{eqnarray}
&&a^1_{1,1}=-0.66605554882019023429631637563968997173,  \nonumber\\
&&a^1_{1,-1}=0.27052090248727273290847449418182448488,\nonumber\\
&&a^1_{-1,1}=3.6002398946839792257416422206042247363,\nonumber\\
&&a^1_{-1,-1}=-0.117428488595883427468489650014095681785,\nonumber\\
&&a^2_{1,1}=1.229132742665024456496131959076675712559,\nonumber\\
&&a^2_{1,-1}=5.796357753327621304524412601597533898065,\nonumber\\
&&a^2_{-1,1}=-2.856445923073834469957280434899450711875,\nonumber\\
&&a^2_{-1,-1}=-1.081767813163632994177953436642495030883,\nonumber\\
&&b_{1,1}=0.078213896418856586401163070522148337413, \nonumber\\
&&b_{1,-1}=1.19199676640685627533532926755144489269, \nonumber\\
&&b_{-1,1}=0.81706609692946543514881835105867033706, \nonumber\\
&&b_{-1,-1}=1. \label{Solvedab2}
\end{eqnarray}

\begin{center}
\begin{minipage}{11cm}
\begin{center} \includegraphics[width=10cm]{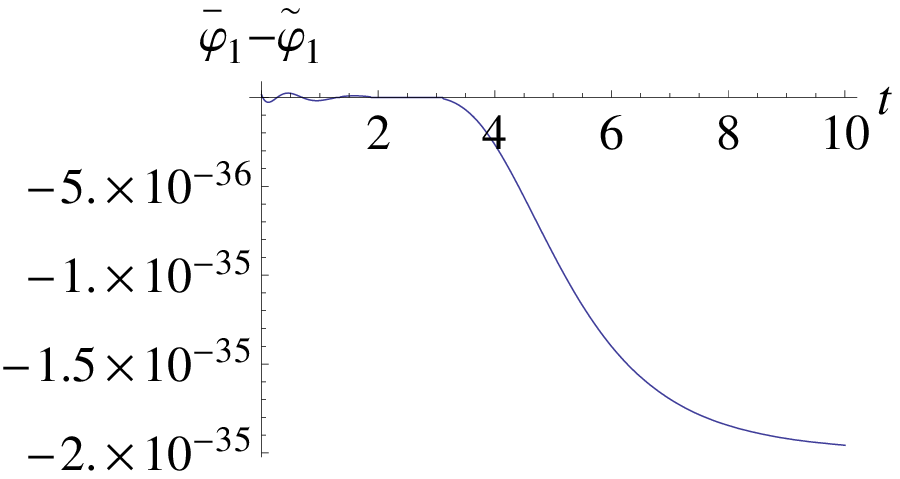} \end{center}
\begin{center}\flushleft
Fig. 3 \quad The difference between  $\bar \varphi_1$ and $\tilde \varphi_1$.
$\bar \varphi_1$ denote $\varphi_1$ with (\ref{Solvedab2}) and
$\tilde \varphi_1$ denote the numerical solution of $\varphi_1$.
\end{center}
\end{minipage}
\end{center}

Fig. 3 shows that the form of $\varphi_1$ must be (\ref{varphiForm2}).

By the numerical values of $\mathsf{f}(t)$, $t \in (0,3)$,
we can determine
\begin{eqnarray}
\mathsf{f}(t)=c_0+c_1 t+\ln \left( \sum \limits_{i,j\in \left\{1,-1 \right\} }
b_{i,j} e^{i \omega_1 t+ j \omega_2 t} \right)^2,
\label{f2}
\end{eqnarray}
and the values of $c_0$ and $c_1$ are
\begin{eqnarray}
&&c_0=-2.254578789614522148435364381123302299, \nonumber\\
&&c_1=3.000000000000000000000000000000000005
\approx \lambda_1+\lambda_2.
\label{Valcs2}
\end{eqnarray}

\begin{center}
\begin{minipage}{11cm}
\begin{center} \includegraphics[width=10cm]{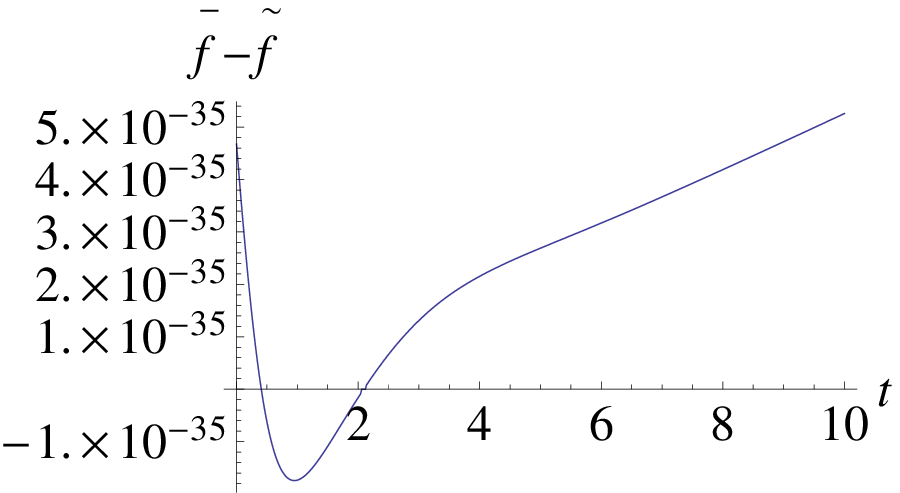} \end{center}
\begin{center}\flushleft
Fig. 4 \quad The difference between $\bar f$ and $\tilde f$.
$\bar f$ denote $\mathsf{f}$ in (\ref{f2}) with (\ref{Valcs2}) and
$\tilde f$ denote the numerical solution of $\mathsf{f}$.
\end{center}
\end{minipage}
\end{center}

Fig.4 shows that the form of $\mathsf{f}$ must be (\ref{f2}).

\subsection{n=3}
\hspace*{0.6cm}
The procedures for $n=3$ is just the same as the one for $n=2$.
All numerical errors  are also far smaller than $10^{-30}$.
We will omit the plots here.

We arbitrarily take the parameters $\lambda_i$, $i=1,2,3$ as
\begin{eqnarray}
\lambda_1=1, \quad \lambda_2=2, \quad \lambda_3=5. \nonumber
\end{eqnarray}
Then we set the following initial values
\begin{eqnarray}
&&\varphi_1(0)=1, \quad \varphi_2(0)=1, \quad \varphi_3(0)=1, \nonumber\\
&&\varphi_1'(0)=1, \quad \varphi_2'(0)=-1, \quad \varphi_3'(0)=-3, \nonumber\\
&&\varphi_1''(0)=-1, \quad \varphi_2''(0)=-1, \quad \varphi_3''(0)=-1,
\nonumber\\
&&\varphi_1'''(0)=-2, \quad \varphi_2'''(0)=-2, \quad \varphi_3'''(0)=-2,
\nonumber\\
&&u(0)=-6, \quad f(0)=0. \label{Initials3}
\end{eqnarray}
We can obtain
\begin{eqnarray}
&&\varphi_1(\infty) \approx \varphi_1(72)=
14.92110270021879759145408479891323291706, \nonumber\\
&&\varphi_2(\infty) \approx \varphi_2(72)=
-41.37915928188619028119192656047612164361 , \nonumber\\
&&\varphi_3(\infty) \approx \varphi_3(72)=
44.53116386683746280518744342554027669437. \label{vpsis3Inf}
\end{eqnarray}

\begin{eqnarray}
M=\left( \begin{array}{ccc}
\lambda_1-\varphi_1(\infty)&-\varphi_1(\infty)&-\varphi_1(\infty)\\
-\varphi_2(\infty) & \lambda_2-\varphi_2(\infty) & -\varphi_2(\infty)\\
-\varphi_3(\infty) &-\varphi_3(\infty) &\lambda_3-\varphi_3(\infty)
\end{array} \right).
\end{eqnarray}

The eigenvalues of $-M$ are
\begin{eqnarray}
&&\Omega_1=6.277038684404317215906811766255007511305, \nonumber\\
&&\Omega_2=2.341721658570986197790603783195012345933, \nonumber\\
&&\Omega_3=1.454346942194766701752186114527368110582.
\end{eqnarray}

Define
\begin{eqnarray}
\omega_1=\frac{1}{2} \Omega_1, \quad \omega_2=\frac{1}{2} \Omega_2,
\quad \omega_3=\frac{1}{2} \Omega_3.
\end{eqnarray}
By the numerical values of $\varphi_1(t), t \in (0, 3)$,
we can determine
\begin{eqnarray}
\varphi_1(t)=\frac{\sum \limits_{i,j,k \in \left\{1,-1 \right\} }
a^1_{i,j,k}e^{i \omega_1 t+ j \omega_2 t+ k \omega_3 t}}
{\sum \limits_{i,j,k \in \left\{1,-1 \right\} }
b_{i,j,k}e^{i \omega_1 t+ j \omega_2 t+ k \omega_3 t}},
\label{varphiForm3}
\end{eqnarray}
where we can assume $b_{-1,-1,-1}=1$.
The calculated $a^1_{i,j,k}$ and $b_{i,j,k}$ are
\begin{eqnarray}
&&a^1_{1,1,1}=0.529262661391713354983735535897686104404,  \nonumber\\
&&a^1_{1,1,-1}= -0.350348759673940084264112387560739906945,\nonumber\\
&&a^1_{1,-1,1}=-8.2675670367386623516544803221517811099,\nonumber\\
&&a^1_{1,-1,-1}=-0.0254209498663860215070864228788457306588,\nonumber\\
&&a^1_{-1,1,1}=-8.2069634233611868635668643982404949776,\nonumber\\
&&a^1_{-1,1,-1}= 0.94894068738511728728577621078955537731,\nonumber\\
&&a^1_{-1,-1,1}=25.8865938214425998112477469386616735198,\nonumber\\
&&a^1_{-1,-1,-1}=-0.80423010558221252574876705199352366617,\nonumber\\
&&b_{1,1,1}=0.035470747170981702373487982709002505528, \nonumber\\
&&b_{1,1,-1}=0.126837598026745943988081084806178876739, \nonumber\\
&&b_{1,-1,1}=1.38001767415062124882858267790796406951, \nonumber\\
&&b_{1,-1,-1}=-0.0229217114935723223611012392981108059801, \nonumber\\
&&b_{-1,1,1}=0.758483231785985808547518495835409278743, \nonumber\\
&&b_{-1,1,-1}=0.473751827832812619094405400960157166769, \nonumber\\
&&b_{-1,-1,1}=5.95862752752346760630497369960292851901,\nonumber\\
&&b_{-1,-1,-1}=1. \label{Solvedab3}
\end{eqnarray}

Similarly the other $a^2_{i,j,k}$ and $a^3_{i,j,k}$
( the coefficients of $\varphi_2(t)$  and $\varphi_3(t)$ )
can also be calculated.
But they do not have much importance, and we will not list them here.

By the numerical values of $\mathsf{f}(t)$,  $t \in (0,3)$,
we can determine
\begin{eqnarray}
\mathsf{f}(t)=
c_0+c_1 t+\ln \left( \sum \limits_{i,j,k \in \left\{1,-1 \right\} }
b_{i,j,k} e^{i \omega_1 t+ j \omega_2 t+ k \omega_3 t} \right)^2,
\label{f3}
\end{eqnarray}
and the values of $c_0$ and $c_1$ are
\begin{eqnarray}
&&c_0=-4.5463675370738389259379741125305027486, \nonumber\\
&&c_1=8=\lambda_1+\lambda_2+\lambda_3.
\label{Valcs3}
\end{eqnarray}

\section{Conclusions}
\hspace*{0.6cm}
By the numerical results in Section 3,
we can  see that for general $n$
\begin{eqnarray}
\mathsf{f}=c_0 +\left( \sum_{i=1}^n \lambda_i \right) t+
\ln \left( \sum_{i_1,i_2,\cdots,i_n \in \{1,-1\} } b_{i_1,i_2,\cdots,i_n}
e^{i_1 \omega_1 t+ i_2 \omega_2 t+\cdots+i_n \omega_n t}
\right)^2.
\label{generalf0}
\end{eqnarray}
It can be shown that $c_0$ and $b_{i_1,i_2,\cdots,i_n}$ are all
dependent on $x$ and that $\omega_i$, $i=1,\cdots, n$ are not
dependent on $x$ and $t$.
Therefore we can write
\begin{eqnarray}
\mathsf{f}=\left( \sum_{i=1}^n \lambda_i \right) t+
\ln \left( \sum_{i_1,i_2,\cdots,i_n \in \{1,-1\} } b_{i_1,i_2,\cdots,i_n}(x)
e^{i_1 \omega_1 t+ i_2 \omega_2 t+\cdots+i_n \omega_n t}
\right)^2,
\label{generalf1}
\end{eqnarray}
where the demand $b_{-1,-1,\cdots,-1}=1$ has been removed.
By (\ref{Evolf}) we can only get $\mathsf{f}_{xt}=v$.
The $\tau$-function of the Ito equation satisfies
$\tau_{xt}=v$ and $\tau_{xx}=\frac{1}{3}u$.
So we can not immediately identify $\tau$ with $\mathsf{f}$.
But we can still show $\tau$ has the same form as (\ref{generalf1})
or we can write
\begin{eqnarray}
\tau=\ln \left( \sum_{i_1,i_2,\cdots,i_n \in \{1,-1\} }
\bar b_{i_1,i_2,\cdots,i_n}(x)
e^{i_1 \omega_1 t+ i_2 \omega_2 t+\cdots+i_n \omega_n t}
\right)^2.
\label{Solvedtau}
\end{eqnarray}
Of course,  $\bar b_{i_1,i_2,\cdots,i_n}(x)$, $i=1,\cdots,n$,
in (\ref{Solvedtau}) are  not arbitrary.
They must satisfy some constraints.
In fact we know there is only one arbitrary function among
all the  $\bar b_{i_1,i_2,\cdots,i_n}(x)$s.

From the numerical results in Section 3,  
we can also see that the eigenfunction $\varphi_j$
must take the form
\begin{eqnarray}
\varphi_j=\frac{\sum\limits_{i_1,i_2,\cdots,i_n \in \{1,-1\} }
\bar a^j_{i_1,i_2,\cdots,i_n}(x)
e^{i_1 \omega_1 t+ i_2 \omega_2 t+\cdots+i_n \omega_n t}}
{\sum\limits_{i_1,i_2,\cdots,i_n \in \{1,-1\} }
\bar b_{i_1,i_2,\cdots,i_n}(x)
e^{i_1 \omega_1 t+ i_2 \omega_2 t+\cdots+i_n \omega_n t}}.
\label{Solvevphis}
\end{eqnarray}

A  discovering of this paper is that the ODE system (\ref{ODEs})
is a  completely integrable one,
which seems to be never seriously studied in the literatures.
Therefore we can conclude that the time evolution for
the multi-solitons  of the Ito equation is integrable.
The dependence on $x$ for the multi-solitons  of the Ito equation is 
Equations (\ref{LaxL}) and  (\ref{ConstrVv}),
which are not completely solved yet.
Recall that the KdV equation with vanishing boundary condition
is completely solved by the multi-soliton solutions of the KdV equation.
Likely we conjecture  that the Ito equation with vanishing boundary condition
is completely solved by the multi-soliton solutions governed by
Equations (\ref{LaxL}),  (\ref{LaxP}),  (\ref{ConstrVv}) and  (\ref{EvolU}).

The method proposed here  provide an  interesting way
from numerical studies to analytical ones. 
We expect this method can also be applied  to some  other nonlinear systems.

\end{document}